\documentclass[conference]{IEEEtran}
\IEEEoverridecommandlockouts

\usepackage{algorithm}
\usepackage{algpseudocode}
\usepackage{multirow}
\usepackage{amsmath}
\usepackage{pifont}
\usepackage{svg}
\usepackage{graphicx}
\usepackage{subcaption}
\usepackage[table]{xcolor}
\usepackage{colortbl}
\usepackage{fontawesome5}
\usepackage{xurl}
\usepackage[hidelinks]{hyperref}

\svgpath{{Figure/}}
\algrenewcommand\algorithmicrequire{\textbf{Input:}}
\algrenewcommand\algorithmicensure{\textbf{Output:}}

\newcommand{\selfcite}[2][blue]{%
  \textcolor{#1}{\cite{#2}}%
}

\newcommand{\selfref}[2][blue]{%
  \textcolor{#1}{\ref{#2}}%
}

\newcommand{\selfurl}[2][blue]{%
  \textcolor{#1}{#2}%
}

\newcommand{\SetCaptionSpacing}[2]{%
    \setlength{\abovecaptionskip}{#1}%
    \setlength{\belowcaptionskip}{#2}%
}

\newcommand{\red}[1]{
    {#1}
}

\newcommand{\Revision}[2]{%
    {#1}
}

\newcommand{\papertitle}{ARMOR}
\newcommand{\subwordTerm}{slice}
\newcommand{\subwordTermBegin}{Slice}

\def\BibTeX{{\rm B\kern-.05em{\sc i\kern-.025em b}\kern-.08em
    T\kern-.1667em\lower.7ex\hbox{E}\kern-.125emX}}
\begin{document}

\title{
\papertitle: Accelerating RTL Simulation by Mitigating the Front-End Bottleneck Using \\ Node Compression
}

\author{
\IEEEauthorblockN{
Jiaping Tang\IEEEauthorrefmark{1}\IEEEauthorrefmark{2}\IEEEauthorrefmark{3},
Jianan Mu\IEEEauthorrefmark{1}\textsuperscript{\faEnvelope[regular]},
Zhiteng Chao\IEEEauthorrefmark{1}\textsuperscript{\faEnvelope[regular]},
Jingzhong Wen\IEEEauthorrefmark{4},
Jing Ye\IEEEauthorrefmark{1}\IEEEauthorrefmark{3},
and Huawei Li\IEEEauthorrefmark{1}\IEEEauthorrefmark{2}\IEEEauthorrefmark{3}
\textsuperscript{\faEnvelope[regular]}
}

\IEEEauthorblockA{
\IEEEauthorrefmark{1}
State Key Laboratory of Processors,
Institute of Computing Technology,
Chinese Academy of Sciences, Beijing, China
}

\IEEEauthorblockA{
\IEEEauthorrefmark{2}
University of Chinese Academy of Sciences, Beijing, China
}

\IEEEauthorblockA{
\IEEEauthorrefmark{3}
CASTEST Co., Ltd., Beijing, China
}

\IEEEauthorblockA{
\IEEEauthorrefmark{4}
Zijing Xinjie Intelligent Technology Co., Ltd., Beijing, China
}

\IEEEauthorblockA{
\{tangjiaping22s, mujianan, chaozhiteng, yejing, lihuawei\}@ict.ac.cn,
jzhwen@siorigin.com
}
}

\maketitle

\begin{abstract}
Register-transfer level~(RTL) simulation is indispensable in chip design.
High-performance simulators typically lower each node in the RTL graph into an instruction sequence. 
Although this per-node lowering enables aggressive compiler optimizations, it dramatically increases the code footprint, severely exceeding instruction cache capacity and causing front-end bottlenecks. 
Our profiling reveals that over 50\% of pipeline stalls are caused by the CPU front-end, becoming a key performance bottleneck in state-of-the-art RTL simulators.
However, reaping the optimization benefits of fully unrolling the RTL graph while simultaneously reducing the code footprint to mitigate front-end bottlenecks remains highly challenging.

In this paper, we propose \papertitle{}, an efficient RTL simulator
designed to alleviate the front-end bottleneck through node compression.
The key idea is to exploit the data parallelism exposed by the unrolling RTL graph and the insufficient bit-space utilization revealed by per-node lowering, leveraging bit-level data parallelism to compress multiple nodes simultaneously, so that a single instruction sequence can serve multiple nodes instead of one per node.
To achieve profitable node compression, 
we first propose a module-aware isomorphic subgraph identification method that leverages structural isomorphism across module instances to systematically identify compression opportunities at the subgraph level.
We then propose an alignment-aware dense packing strategy that groups nodes into packs according to dataflow dependencies while preserving data reuse, complemented by greedy merging strategies to enhance bit-space utilization.
Finally, we implement a unified bit-level parallelism scheme to support bit-level parallel execution of compressed nodes.
Experimental results show that \papertitle{} compresses 57\% of nodes on average, effectively mitigating front-end bottlenecks, and achieving 1.6$\times$ speedup on CPU designs and 2.7$\times$ speedup on AI accelerators compared to state-of-the-art simulators.
\end{abstract}


\section{Introduction}

RTL simulators are fundamental software tools in the digital circuit design flow, frequently employed during both design and debugging. Improving RTL simulation performance significantly enhances chip development efficiency and has long been a key topic of interest in both academia and industry~\selfcite{micro_beamer, GSIM, verilator, repcut, dedup, Xcelium,synopsys_vcs, RTeAAL_ASPLOS_2026}.
Despite the emergence of various hardware-assisted simulation and emulation platforms~\selfcite{elsabbagh2023accelerating, emami2023manticore, FireAxe,firesim,cadence_emulation,synopsys_emulation}, the mainstream approach for RTL simulation remains software-based execution on general-purpose CPUs, especially in the early stages of design iteration, due to their flexibility and ease of debugging~\selfcite{micro_beamer}.


RTL simulation can be viewed as a topological evaluation of a circuit dataflow graph called \textit{RTL graph}, where each node is computed in dependency order to generate the circuit state for the next cycle. 
To accelerate evaluation of RTL graphs, prior works~\selfcite{GSIM,verilator,repcut, dedup, RTeAAL_ASPLOS_2026, essent, zhou2023khronos,parendi} have explored a range of optimizations, including reducing redundant computation, optimizing memory access patterns, and minimizing inter-core synchronization overhead.
Despite achieving significant performance gains, these approaches adopt a \textit{node-granular execution model}, where each node is lowered into instructions and executed individually. 
This execution model leads to long instruction streams, resulting in increased code footprint.


\begin{figure}
    \SetCaptionSpacing{0.10cm}{0.10cm}
    \centering
    \includegraphics[width=0.9\linewidth]{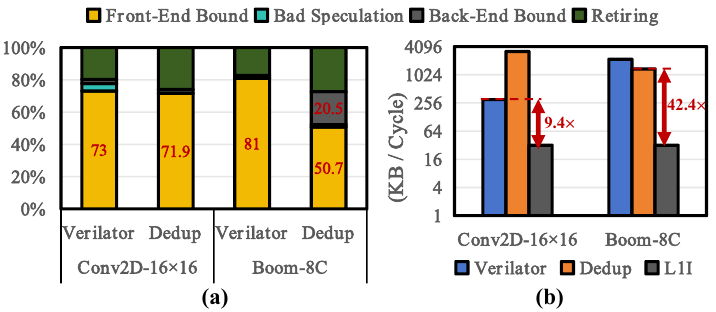}
    \caption{
        Profiling data for Conv2D~\selfcite{tensorlib} and Boom designs~\selfcite{boom_core} are collected using Intel VTune~\selfcite{intel_vtune} on the Platinum 8338 processors.
        (a): Micro-architectural profiling results of the Verilator~\selfcite{verilator} and Dedup~\selfcite{dedup}.
        (b): Instruction working set size per cycle of Verilator and Dedup.
    }
    \label{fig:intro}
\end{figure}

RTL simulation on CPUs suffers from a significant front-end bottleneck. 
As shown in Figure~\selfref{fig:intro}~(a), for Verilator~\selfcite{verilator}, over 70\% of CPU pipeline slots go unoccupied due to insufficient micro-ops supply from the front-end.
Our analysis identifies L1-I cache misses as the principal cause: 
RTL graphs often consist of numerous nodes, each lowered into an instruction sequence, making the instruction working set excessively large.
As shown in Figure~\selfref{fig:intro}~(b), the working set per cycle can grow to hundreds or thousands of kilobytes, vastly exceeding the 32 KB L1 instruction cache capacity by more than an order of magnitude, causing frequent cache misses.
Besides, the working set grows with design size, exacerbating the bottleneck.
\textbf{Therefore, a key to accelerating CPU-based RTL simulation is to reduce the code footprint, thereby alleviating pressure on the CPU front-end.}

\begin{figure}
    \SetCaptionSpacing{0.10cm}{0.10cm}
    \centering
    \includegraphics[width=\linewidth]{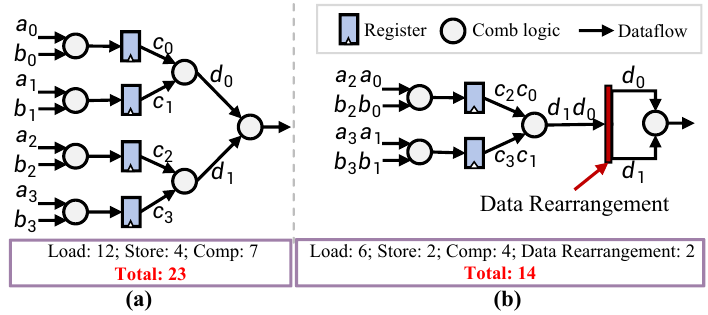}
    \caption{
        Example of a 2-stage pipeline.
        RTL simulation without node compression~(a) and with node compression~(b).
    }
    \label{intro_example}
\end{figure}

Structured or reusable representations of the RTL graph appear to be a natural way to reduce code footprint. 
For example, loop-based representations roll repeated computation into compact kernels~\selfcite{RTeAAL_ASPLOS_2026}, and code-reuse approaches extract common RTL regions into reusable code~\selfcite{dedup}. 
However, both approaches reduce code footprint by sacrificing the static specialization provided by fully unrolled code, such as hard-coded signal accesses and compile-time-visible dataflow~\selfcite{dedup, RTeAAL_ASPLOS_2026, ASPLOS_2022_Theodoridis}. 
This loss of static specialization turns compile-time-resolved signal accesses into runtime-resolved memory accesses, thereby exacerbating the back-end bottleneck, as shown in Figure~\selfref{fig:intro}~(a) when Dedup simulates Boom-8C.
Moreover, it also limits the compiler’s ability to perform aggressive optimizations~\selfcite{RTeAAL_ASPLOS_2026}.
Overall, fully unrolled code usually provides strong performance advantages over rolled representations by enabling static hard-coding and compiler optimization opportunities~\selfcite{RTeAAL_ASPLOS_2026}, and therefore remains the mainstream representation for compiled RTL simulation~\selfcite{GSIM,verilator,repcut, essent, zhou2023khronos,parendi}.
\textbf{Therefore, the key challenge in optimizing RTL simulation lies in simultaneously achieving two goals: keeping the RTL graph fully unrolled to reap the benefits of static specialization, while reducing the code footprint caused by full unrolling to avoid front-end bottlenecks.}


Although unrolling the RTL graph increases code footprint, it also exposes abundant data parallelism, creating opportunities for optimizing code footprint.
\textbf{Our key insight is to exploit this parallelism via node compression: multiple isomorphic nodes are evaluated simultaneously using a single instruction sequence instead of one per node.}
This approach preserves the fully unrolled RTL graph while avoiding the code footprint growth caused by unrolling, thereby reducing the instruction working set and alleviating front-end bottlenecks.
As illustrated in Figure~\selfref{intro_example} with a simple yet intuitive and powerful example, without node compression shown in Figure~\selfref{intro_example}~(a), each simulation cycle requires 23 operations.
In contrast, with node compression shown in Figure~\selfref{intro_example}~(b), each cycle requires only 14 operations.
This reduction not only reduces the code footprint but also decreases the computational workload, thereby improving simulation performance.

In this paper, we propose \papertitle{}\footnote{\selfurl{https://github.com/jptang99-star/ARMOR.git}}, an efficient RTL simulator that alleviates front-end bottlenecks through node compression. 
Our approach is motivated by two key observations. 
First, machine words are significantly underutilized, leading to bit-level waste. 
Second, unrolled RTL graphs expose abundant parallelism among isomorphic nodes, enabling multiple nodes to be executed simultaneously. 
These observations suggest an opportunity to increase the per-instruction workload by compressing multiple nodes into a single instruction sequence.
To realize this idea, we design a node compression framework that systematically exploits such opportunities. 
We first identify isomorphic subgraphs using a module-aware approach to expose compression opportunities at the subgraph level. 
We then develop an alignment-aware packing strategy that maximizes the number of nodes mapped into a machine word while preserving data reuse. 
Finally, we implement a unified bit-level parallel execution scheme that supports common operations in the RTL graph and efficiently executes compressed nodes.
We evaluate \papertitle{} on a set of AI accelerators and CPU designs. 
Experimental results show that \papertitle{} compresses 57\% of nodes on average, and achieves 2.7$\times$ speedup on AI accelerator designs and 1.6$\times$ speedup on CPU designs compared to state-of-the-art simulators.

This paper makes the following contributions:
\begin{itemize}
    \item We propose node compression, implemented via bit-level data parallelism, which exploits both the inherent parallelism of isomorphic nodes in the RTL graph and the underutilization of bit-space to reduce code footprint and mitigate front-end bottlenecks in RTL simulation.
    \item We propose a module-aware isomorphic subgraph identification method, design an alignment-aware packing strategy, and implement a unified bit-level parallel execution scheme, making node compression practically profitable.
    \item We demonstrate that \papertitle{} is effective and scalable in mitigating front-end bottlenecks and delivering performance improvements across diverse RTL designs.
\end{itemize}

\section{Background}

\subsection{RTL Simulation Overview}
\begin{figure}[htbp]
    \SetCaptionSpacing{0.10cm}{0.10cm}
    \centering
    \includegraphics[ width=\linewidth]{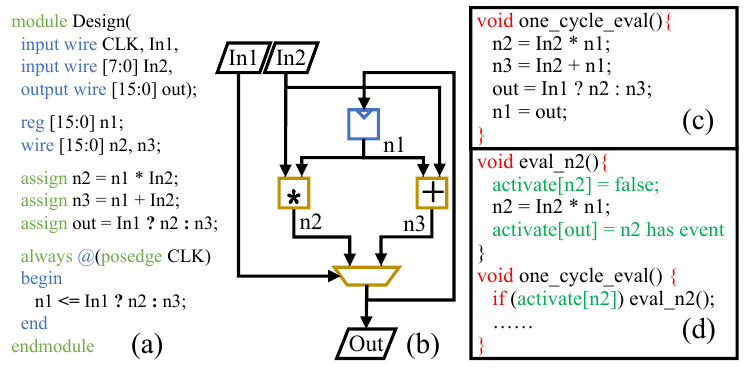}
    \caption{
        Example of RTL code~(a) and its corresponding dataflow graph~(b).
        Compiled full-cycle simulator~(c) and event-driven simulator~(d) that only shows the \textit{eval} function of node \textit{n2}.
        Event management is highlighted in green.
        The \textit{one\_cycle\_eval} function simulates one cycle per execution.
    }
    \label{background_sim}
\end{figure}

RTL simulation converts hardware designs described in RTL into a dataflow graph, where nodes represent components~(e.g., adders, registers) and edges represent wires connecting them~\selfcite{elsabbagh2023accelerating, RIROS, ERASER}, as shown in Figures~\selfref{background_sim}~(a) and (b). This graph is then translated into a software program, as illustrated in Figures~\selfref{background_sim}~(c) and (d), compiled into a simulator~\selfcite{elsabbagh2023accelerating}.
There are two main types of RTL simulation: \textit{event-driven} and \textit{full-cycle or cycle-based}.
In event-driven simulation, each signal change triggers events, propagating updates until no further events occur. This minimizes computation but incurs event management overheads~(green-highlighted in Figure~\selfref{background_sim}~(d)).
Full-cycle simulation, also called \textit{oblivious} simulation, evaluates the entire design every clock cycle, even when inputs remain unchanged, leading to redundant computations but avoiding scheduling overhead.
Both types of simulation have their respective application scenarios: event-driven suits low-activity designs like CPUs, while full-cycle suits high-activity designs like AI accelerators.

Besides, event-driven simulation and full-cycle simulation can leverage compilation techniques, with compiled simulators outperforming interpreted ones by nearly 100$\times$~\selfcite{verilator, Xcelium}. Figures~\selfref{background_sim}~(c) and (d) show compiled implementations for full-cycle and event-driven simulations, respectively. Compiled simulators unroll the entire dataflow graph, making memory access patterns and computations statically known at compile time, enabling aggressive compiler optimizations. However, this unrolling can overwhelm the processor’s front-end due to the large instruction working set.

\subsection{Front-end Bottleneck}\label{bg_frontend}

The processor \textit{front-end}
is responsible for fetching, decoding, and delivering micro-ops to the rest of the processor pipeline, called the \textit{back-end}~\selfcite{frontend_leaky, frontend_topdown}.
Front-end bound describes a performance bottleneck where the processor stalls due to limitations in the front-end stages, such as instruction-cache misses and insufficient decode bandwidth.
When front-end bound dominates, the instruction supply rate is insufficient to keep the back-end fully utilized, thus resulting in poor performance~\selfcite{frontend_leaky, frontend_weeding, frontend_asmdb, frontend_clear, frontend_softsku, frontend_profiling}. 
Some works~\selfcite{frontend_weeding,  frontend_cache_replace,frontend_ispy, ICARUS_Kalbande_ASPLOS2026} tailored for the front-end bottleneck have proposed optimization strategies at the hardware level, such as improving instruction cache replacement and instruction prefetching. 
However, due to the unique characteristics of RTL simulation tasks, such as the presence of an extremely large instruction working set and long reuse interval, these optimizations struggle to deliver the expected performance gains. 
Moreover, compared to software-level approaches, hardware-level optimizations typically require modifying the processor micro-architecture, which is costly, inflexible, and unable to benefit the vast number of already-deployed CPUs.
Collectively,  these factors underscore the urgent need for targeted solutions that address front-end inefficiencies in RTL simulation, which we focus on in this paper.


\section{Motivation \& Challenge}

\subsection{Inefficiency of Node-Granular Execution}

\begin{figure}[htbp]
    \SetCaptionSpacing{0.10cm}{0.10cm}
    \centering
    \includegraphics[width=\linewidth]{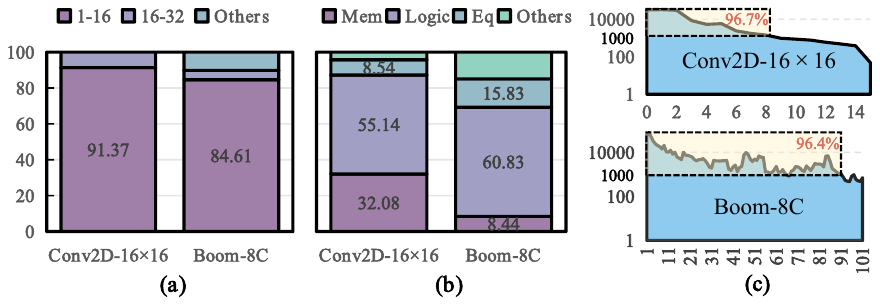}
    \caption{
        Distribution of (a) Node bit-widths, (b) Node types, and (c) Node counts across topology levels in the RTL graph.
        For the Boom-8C design, the first 10\% of topology levels are reserved for better visualization.
    }
    \label{fig:motivation1}
\end{figure}

{We perform a systematic analysis of RTL simulation task characteristics and hardware resource utilization efficiency, yielding the following key observations:}

\textbf{Observation 1: Bit-space inefficiency.}
Our analysis of the bit-width distribution of various nodes in RTL graphs, as shown in Figure \selfref{fig:motivation1}~(a), reveals that over 80\% of nodes operate on fewer than 16 bits. 
In contrast, modern CPUs are designed to handle 64-bit operations efficiently.
Prior works~\selfcite{GSIM,verilator,repcut, dedup, essent, zhou2023khronos,parendi} extend these low-bitwidth operands to standard widths (e.g., 16 or 32 bits), resulting in severe under-utilization of the available bit-space and inefficient use of computational resources, which we term \textit{bit-space inefficiency}.
We evaluate a set of benchmarks, as detailed in Section~\selfref{sec:bit_utilization}, and observe that certain designs exhibit critically low bit-space utilization, with utilization below 20\%.

\begin{figure*}
  \centering

  \begin{minipage}{0.25\linewidth}
    \centering
    \includegraphics[width=\linewidth]{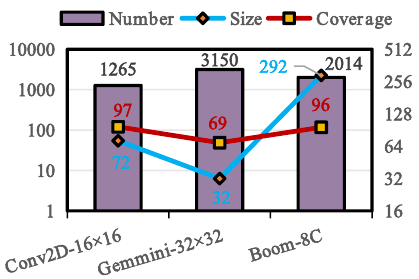}
    \caption{
    \Revision{Number, Size, and Coverage denote the number of isomorphic subgraph instances of any structure, their average size, and the percentage of nodes covered by these subgraphs.}{A}
    }
    \label{fig:sub_graph}
  \end{minipage}
  \hfill
  \begin{minipage}{0.72\linewidth}
    \centering
    \includegraphics[width=\linewidth]{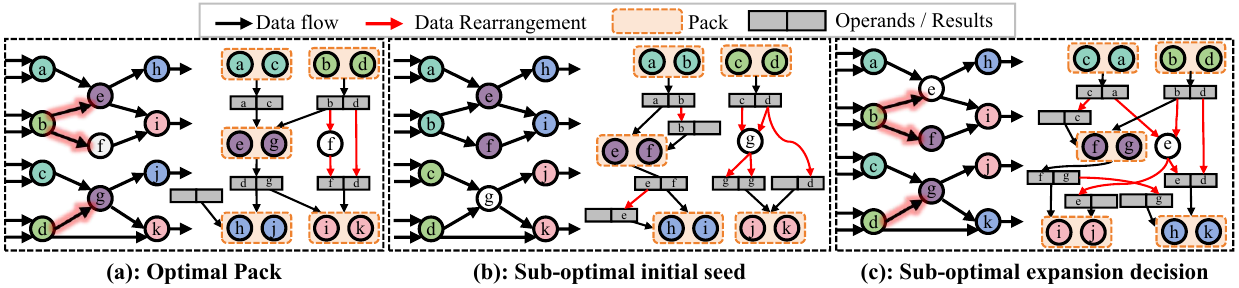}
    \caption{
        In this illustrative example, nodes at the same topological level are isomorphic, and same-colored nodes form a pack.
        Edges highlighted in red indicate the expansion from \{\textit{b,d}\} to next-level candidates.
        (a): An ideal packing result requires only three data rearrangements.
        (b): A sub-optimal initial seed selection chooses \{\textit{a,b}\} and \{\textit{c,d}\} as seed packs.
        (c): A sub-optimal expansion decision expands from \{\textit{b,d}\} to \{\textit{f,g}\} instead of \{\textit{e,g}\}.
        Both sub-optimal choices result in seven data rearrangements.
    }
    \label{fig:challenge}
  \end{minipage}

\end{figure*}
\textbf{Observation 2: Abundant data-level parallelism.}
The RTL graph exhibits a highly skewed node-type distribution, dominated by a few node types. 
As shown in Figure~\selfref{fig:motivation1}~(b), memory-access nodes~(e.g., registers declared in RTL designs), bitwise logical operation nodes~(e.g., AND, OR, XOR, and NOT), and Eq operation nodes account for approximately 32.08\%, 55.14\%, and 8.54\% of all nodes in the Conv2D-$16\times16$ design, respectively.
This indicates the prevalence of \textit{isomorphic nodes} that share the identical operation type but differ only in operands.
Furthermore, the RTL graph also exhibits a pronounced skewed topology, with nodes highly concentrated in a small number of topological levels. 
Topological analysis, as shown in Figure~\selfref{fig:motivation1}~(c), reveals that nodes are highly concentrated in the first few levels, with levels containing over 1,000 nodes accounting for more than 96\% of all nodes in the RTL graph.
Critically, nodes within the same topological level can be computed independently.
Taken together, these characteristics exhibited by RTL graphs indicate that RTL simulation exhibits abundant data-level parallelism, with large groups of isomorphic nodes amenable to parallel evaluation.


\textbf{Overall, these observations indicate a granularity mismatch between RTL node evaluation and host-machine execution. Conventional node-granular execution underutilizes the bit-space within host machine words and leaves abundant independent node-level evaluation unexploited.}

\subsection{Motivation: From Isolated Compression to Subgraph Reuse}

The above observations reveal opportunities for \textit{compressed execution}, where multiple narrow RTL nodes are mapped to different bit regions of the same machine word and are evaluated together.
However, compressing nodes based only on locally identical operators is insufficient.
If the surrounding dataflow offers limited opportunities for reusing compressed results, the output of one compressed node group may not be directly consumed by downstream compressed nodes, introducing \textit{data rearrangement overhead}, i.e., extra operations to remap packed bits for downstream use.
Therefore, effective compressed execution requires repeated computation structures whose producer-consumer relationships are also preserved.
To characterize whether RTL graphs provide such opportunities, we further profile repeated subgraphs in RTL designs.
As shown in Figure~\selfref{fig:sub_graph}, nearly 70\% of RTL nodes are covered by structurally repeated subgraphs, referred to as \textit{isomorphic subgraphs}.
In these isomorphic subgraphs, nodes at the same structural positions have consistent producer-consumer relationships, providing opportunities for reusing compressed results.
This indicates that compression opportunities in RTL graphs are not limited to isolated nodes, but can propagate along producer-consumer dependencies across isomorphic subgraph instances.

To this end, these results provide a structural basis for \textit{node compression}.
Instead of lowering each RTL node into a separate instruction sequence, node compression aggregates multiple parallelizable RTL nodes into different bit positions of the same host word and lowers them collectively into a shared instruction sequence.
This reduces both the generated instruction working set and the number of executed node-evaluation operations, thereby mitigating the front-end bottleneck of compiled RTL simulation.


\subsection{Challenges in Node Compression}

Despite these opportunities, realizing node compression is non-trivial.
A complete compressed execution flow must answer three questions.
First, ARMOR must discover node groups that preserve reusable producer-consumer structures, rather than relying on isolated local compression.
Second, it must place nodes into machine words while balancing bit-space utilization with reusable downstream bit layouts.
Third, it must generate correct word-level execution for operations that introduce dependencies across compressed bit regions.
These requirements lead to the following challenges.

\Revision{

\textbf{Challenge 1: Subgraph-level correspondence in Pack Discovery.}\label{sec:challenge1}
The first challenge is to discover groups of RTL nodes with consistent correspondence across producer-consumer dependencies.
A natural question is whether existing superword-level parallelism (SLP) vectorization can be directly applied.
Existing general-purpose discovery approaches~\selfcite{PLDI_2000_Larsen, ASPLOS_2021_Chen, Shahbahrami2006, MICRO_2017_Huh, MICRO_2010_Barik, PACT_2017_SuperGraph, CGO_2015_PSLP} used in SLP typically start from seed packs and expand them along \textit{use-def} or \textit{def-use} chains~\selfcite{PLDI_2000_Larsen}.
This dependency-driven expansion can preserve reuse between producer and consumer packs, but the resulting correspondence is highly dependent on local decisions~\selfcite{ASPLOS_2012_Park}, including the initial seed selection and subsequent expansion choices.
Different local choices may induce different correspondences among downstream nodes, leading to different final packs and data-rearrangement costs.
Figure~\selfref{fig:challenge} contrasts the ideal case with two typical failure cases.
Specifically, Figure~\selfref{fig:challenge}~(a) shows an ideal packing result, resulting in only three data rearrangements.
In contrast, Figure~\selfref{fig:challenge}~(b) shows a sub-optimal initial seed selection: choosing \{\textit{a,b}\} and \{\textit{c,d}\} as seed packs is locally reasonable, but it induces an unfavorable downstream correspondence and increases the number of data rearrangements to seven.
Figure~\selfref{fig:challenge}~(c) shows another failure mode: even with a good initial seed, a sub-optimal expansion decision, e.g., expanding from \{\textit{b,d}\} to \{\textit{f,g}\} instead of \{\textit{e,g}\}, propagates an unfavorable correspondence and also results in seven data rearrangements.
However, if pack discovery can recognize \{\textit{a,b,e,h}\} and \{\textit{c,d,g,j}\} as two isomorphic subgraphs in advance in Figure~\selfref{fig:challenge}~(a), it can directly derive the desired correspondence, such as \{\textit{a,c}\}, \{\textit{b,d}\}, \{\textit{e,g}\}, and \{\textit{h,j}\}, thereby achieving an ideal packing result.
Such subgraph-level correspondence directly avoids relying solely on local packability and provides stable and reliable guidance for both seed selection and pack expansion.

Obtaining such subgraph-level correspondence is difficult because it hinges on identifying isomorphic subgraphs. Thus, the key challenge is to recover this correspondence without resorting to a generic exhaustive subgraph-isomorphism search over the entire RTL graph.
}{A, C}

\textbf{Challenge 2: Conflict between bit-space utilization and layout reuse.}\label{sec:challenge2}
After pack discovery identifies candidate nodes for compression, node packing must balance \textit{bit-space utilization} and \textit{layout reuse}.
Here, layout reuse means that the bit positions produced by one pack match the bit positions expected by its downstream consumer pack.
For example, if a producer pack \{\textit{a,c}\} feeds a consumer pack \{\textit{b,d}\}, using the same bit ordering allows the values of \textit{a} and \textit{c} to be directly consumed by \textit{b} and \textit{d}, respectively, without data rearrangement.
However, packing nodes to preserve such producer-consumer layouts may leave bits within a machine word underutilized, lowering overall utilization.
Conversely, packing strategies that aggressively maximize bit utilization may combine nodes with incompatible producer or consumer layouts, introducing frequent data rearrangement.
As these objectives are inherently conflicting, achieving high bit-space utilization while maintaining reusable bit layouts constitutes a challenge in node packing.


\textbf{Challenge 3: Incompatibility between bit-level parallel execution and cross-\subwordTerm{} dependent operations.}\label{sec:challenge3}
Node compression maps multiple nodes onto a single machine word to enable bit-level parallel execution, where each node is allocated a contiguous group of bits, referred to as a bit \textit{slice}.
However, this technique only applies to \subwordTerm{}-independent operations, such as logical operations; any cross-\subwordTerm{} dependencies would prevent correct execution.
For example, for equality comparisons, directly applying a scalar equality comparison to two packed operands produces
\(
R = \bigwedge_{l=0}^{L-1} \text{Eq}(X_l, Y_l),
\)
where \(X_l\) and \(Y_l\) are the operands of \subwordTerm{} \(l\), and \(L\) is the number of \subwordTerm s. 
Ideally, each \subwordTerm{} would produce an independent 0/1 result, but the cross-\subwordTerm{} dependency forces aggregation into a single output. 
Consequently, supporting cross-\subwordTerm{} dependent operations while preserving bit-level parallelism remains a challenge in the compressed execution stage.

To overcome these challenges, we propose the \papertitle{} framework, which is introduced in the following sections.

\section{Framework of \papertitle{}}

\begin{figure}[htbp]
    \SetCaptionSpacing{0.10cm}{0.10cm}
    \centering
    \includegraphics[width=\linewidth]{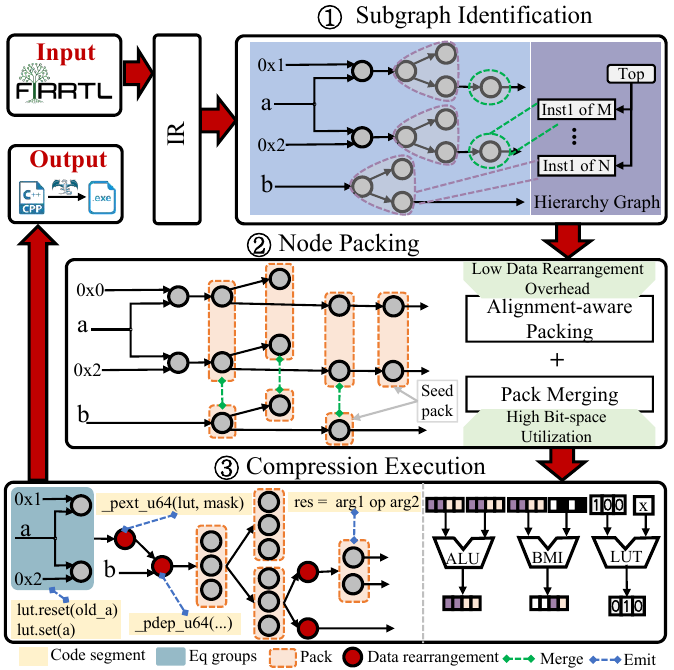}
    \caption{
        An overview of \papertitle{}.
    }
    \label{fig:framework}
\end{figure}

\subsection{\Revision{Overview}{A}}
As illustrated in Figure~\selfref{fig:framework}, the \papertitle{} framework takes FIRRTL~\selfcite{firrtl_sepc} files as input, applies bit-level data parallelism to compress nodes, and ultimately lowers it to a C++ file that can be compiled into a cycle-based RTL simulator.

\red{
The node-compression pipeline consists of three stages.
First, ARMOR identifies isomorphic subgraphs from the RTL graph.
Second, guided by the identified isomorphic subgraphs, ARMOR packs parallelizable isomorphic nodes into machine-word-sized packs.
Finally, ARMOR lowers these packs into operation-specific C++ implementations.
In the generated C++ code, slice-independent operations are emitted as scalar bitwise operations, while cross-slice or layout-manipulation operations that require bit extraction/deposition explicitly use BMI2 intrinsics such as \texttt{\_pext\_u64} and \texttt{\_pdep\_u64}, which are then lowered by the C++ compiler to target \texttt{PEXT}/\texttt{PDEP} instructions on BMI2-supported platforms.
}

The following subsections describe each stage in detail and explain how ARMOR addresses the corresponding challenges: efficient subgraph-level discovery~(Section \selfref{sec:subgraph_identification}), alignment-aware dense packing~(Section \selfref{sec:packing}), and correct compressed execution for diverse RTL operations~(Section \selfref{sec:compression_execution}).

\subsection{Module-Aware Isomorphic Subgraph Identification \& Alignment}\label{sec:subgraph_identification}

Identifying isomorphic subgraphs is a fundamental prerequisite for effective node compression, as it provides a subgraph-level compression perspective. 
Without such a perspective, compressible nodes are discovered in a scattered and short-sighted manner~\selfcite{PLDI_2000_Larsen, ASPLOS_2021_Chen}, preventing coordinated packing across structurally equivalent regions and leading to increased data rearrangement overhead. 
However, directly identifying isomorphic subgraphs on an RTL graph is computationally impractical, as RTL graphs are large in scale and subgraph isomorphism mining is an NP-hard problem~\selfcite{graph_iso1,graph_iso2}.

Fortunately, RTL graphs are not arbitrary structures but are structured by module instantiation, especially in highly parallel architectures with extensive module reuse~\selfcite{high_parallel_archi1,high_parallel_archi2,high_parallel_archi3, HPCA_2026_Yu}.
Specifically, each module instance corresponds to a subgraph in the RTL graph, and multiple instances of the same module naturally exhibit structural isomorphism. 
This structural regularity allows us to bypass general isomorphic subgraph mining and instead leverage module hierarchy as a deterministic mechanism for identifying isomorphic subgraphs.
Based on this observation, we propose a \textit{module-aware isomorphic subgraph identification \& alignment} approach that leverages module structure to deterministically identify and group structurally equivalent nodes across instances, enabling coordinated compression at the subgraph level.

To jointly capture structural information and data dependencies, we construct a hierarchical graph and a flat dataflow graph for the RTL design. 
The flat dataflow graph $G_f = (V, E_f)$ models computation-level dependencies, while the hierarchical graph $G_h = (M, E_h)$ represents module instances and their containment relationships. Each node $v \in V$ is associated with a module instance $m \in M$ via a mapping $\phi: V \rightarrow M$, preserving module context.

We define \textit{compression domains} $\mathcal{D}$ as proxies for isomorphic subgraphs, grouping structurally equivalent nodes across module instances and serving as the basis for subsequent packing. The isomorphic subgraph identification \& alignment algorithm~(Algorithm~\selfref{alg:module-aware}) proceeds as follows. 
\begin{itemize}
    \item First, module instances in $G_h$ are grouped by module type. For each module type $T$, we collect all instances $I_T=\{I_1, I_2, \dots, I_k\}$.
    
    \item Second, for each instance $I_i$, we extract the corresponding node set
    $ S_i = \{ v \in V \mid \phi(v) = I_i \} $ from the flat dataflow graph $G_f$. 

    \item Finally, we align structurally equivalent nodes and collect them, forming $\mathcal{D}$. 
    Instead of performing general subgraph isomorphism testing, we adopt a name-based alignment strategy. Since node names within a module definition are unique, nodes across different instances can be directly aligned by matching their names. For each node name in module type $T$, we aggregate nodes with the same name across all instances to form a compression domain $D_n$, and collect all such domains to obtain $\mathcal{D}$.

\end{itemize}



This algorithm bypasses costly subgraph isomorphism mining and testing, reducing the problem from combinatorial matching to simple linear-time grouping. 
The complexity is linear in the number of module instances and the number of nodes per module, making it highly practical.
Overall, our module-aware approach enables efficient identification and alignment of isomorphic subgraphs, directly addressing the challenge of identifying isomorphic structures, and finally providing subgraph-level coordination for node compression in the RTL graph.

\begin{algorithm}[t]
\caption{Module-Aware Subgraph Identification \& Alignment}
\label{alg:module-aware}

\begin{algorithmic}[1]
\Require Flat dataflow graph $G_f = (V, E_f)$
\Require Hierarchical graph $G_h = (M, E_h)$
\Require Mapping $\phi: V \rightarrow M$
\Ensure Compression domains $\mathcal{D}$

\State $\mathcal{D} \gets \emptyset$

\Statex \textcolor{blue}{// Isomorphic subgraph identification}

\State Group module instances in $M$ by module type

\ForAll{module type $T$}

    \State $I_T \gets$ instances of type $T$

    \State Extract subgraphs 
           $S_i = \{ v \in V \mid \phi(v) = I_i \}$ 
           for each $I_i \in I_T$



    \Statex \textcolor{blue}{// Isomorphic subgraph alignment}
    \State Let $\mathcal{N}_T$ be the set of node names 
           defined in module type $T$

    \ForAll{node name $n \in \mathcal{N}_T$}
        \State $D_n \gets \emptyset$ \Comment{$D_n$: one group of structurally aligned nodes }

        \ForAll{instance $I_i \in I_T$}
            \State $v \gets$ node in $S_i$ with name $n$
            \State Add $v$ to $D_n$
        \EndFor

        \State Add $D_n$ to $\mathcal{D}$
    \EndFor
    
\EndFor

\State \Return $\mathcal{D}$

\end{algorithmic}
\end{algorithm}

\subsection{Alignment-Aware Dense Node Packing}\label{sec:packing}

\Revision{
While compression domains contain structurally equivalent isomorphic nodes across isomorphic subgraphs, they do not directly map to hardware-executable units.
To become executable units, nodes must be organized into \emph{packs} under several constraints.
First, each pack must satisfy the hardware width constraint, i.e., the total bit-width of packed nodes cannot exceed the machine-word width.
Second, nodes within a pack must be dependency-free so that they can be executed in parallel.
In addition to these basic legality constraints, pack construction must preserve structural alignment across isomorphic subgraphs.
This alignment enables subgraph-level data reuse between packs and reduces unnecessary data rearrangement.
}{A}

A fundamental challenge in node packing is the inherent conflict between \textit{bit-space utilization} and \textit{data reuse efficiency}. 
On the one hand, maximizing bit-space utilization requires densely packing nodes to fully utilize machine words. 
On the other hand, high data reuse relies on preserving producer–consumer alignment so that data produced by upstream packs can be directly consumed by downstream packs. 
Packing decisions that focus solely on bit-width often break this alignment, forcing data to be rearranged across packs and introducing data rearrangement overhead, while focusing on data reuse alone may lead to underutilization of bit-space.
To address this challenge, we propose \textit{alignment-aware dense node packing}, a heuristic strategy that explicitly balances this trade-off by jointly optimizing data reuse and bit-space utilization.



First, to improve data reuse, we adopt an \textit{alignment-aware expansion} strategy that expands packs along dataflow dependencies while maintaining structural alignment, enabling both intra- and inter-subgraph data reuse.
To maintain structural alignment during expansion, we expand packs along the dependencies of compression domains in a coordinated manner.
Specifically, during expansion, we ensure that all nodes within each newly expanded pack belong to the same compression domain, thereby preserving producer–consumer alignment across packs and enabling direct data reuse with minimal data rearrangement.

Second, to improve bit-space utilization, we introduce two complementary strategies. 
In the initial pack construction phase, we first select a \textit{maximal feasible group} of nodes that fills a machine word as densely as possible subject to hardware width constraints, thereby providing high initial utilization. 
\Revision{
Specifically, for a machine-word width $W$ and a node width $w$ (the maximum bitwidth among the node's operands and result), each maximal feasible group selects up to $\lfloor W / w \rfloor$ remaining nodes from the same domain.
}{B}
After alignment-aware expansion, we further perform \textit{intra-domain pack merging}, where multiple packs within the same compression domain are merged whenever their combined bit-width does not exceed the machine word size. 
Since these packs remain within the same compression domain and preserve structural alignment, such merging rarely induces data rearrangement.


\begin{algorithm}[t]
\caption{Alignment-Aware Dense Node Packing}
\label{alg:packing}
\begin{algorithmic}[1]

\Require Compression domains $\mathcal{D}$
\Require Machine word size $W$
\Require Bit-width function $w(\cdot)$
\Require Consumer function $cons(\cdot)$
\Ensure Pack set


\Statex \textcolor{blue}{// Partition compression domain}
\State $\mathcal{S} \gets \Call{PartitionByTopologicalLevel}{\mathcal{D}}$

\Statex \textcolor{blue}{// Each node $N \in G$ stores $N.domain$ and $N.packs$}
\State $G \gets \Call{BuildDependencyGraph}{\mathcal{S}}$

\Statex \textcolor{blue}{// Alignment-aware expansion}
\ForAll{nodes $N \in G$ in topological order}
    \While{$N.domain \neq \emptyset$}
        \State $P_N \gets \Call{SelectMaxFeasibleGroup}{N, W}$ 
        \State frontier $\gets \{ (N, P_N) \}$
        \State update $N$ using $P_N$
        \While{frontier $\neq \emptyset$} \Comment{Try to expand}
            \State pop $(F, P_F)$ from frontier \Comment{Seed pack $P_F$}
            \ForAll{successor nodes $S$ of $F$ in $G$}
                \State $P_S \gets cons(P_F) \cap S.domain$ \Comment{New pack}
                \If{$w(P_S) \le W \;and\; |P_S| > 1$} 
                    \State add $(S, P_S)$ to frontier 
                    \State update $S$ using $P_S$
                \EndIf
            \EndFor
        \EndWhile

    \EndWhile
\EndFor

\Statex \textcolor{blue}{// Intra-domain merging} \Comment{Dense}
\ForAll{ nodes $N \in G$}
    \While{$\exists P_i, P_j \in N.packs$ s.t. $w(P_i \cup P_j) \leq W$}
        \State $P_{\text{new}} \gets P_i \cup P_j$
        \State $N.packs \gets (N.packs \setminus \{P_i, P_j\}) \cup \{P_{\text{new}}\}$
    \EndWhile
\EndFor

\State \Return $\bigcup_{N \in G} N.packs$ \Comment{Collect packs across all nodes}

\end{algorithmic}
\end{algorithm}

Based on these ideas, we propose the alignment-aware dense node packing algorithm (Algorithm~\selfref{alg:packing}).
\begin{itemize}
    \item \Revision{First, we partition each compression domain according to the topological levels of the RTL graph, grouping nodes at the same level into a sub-compression domain. This ensures that nodes subsequently grouped into the same pack are dependency-free and can therefore be executed in parallel.}{B} 
    
    \item Second, we construct a dependency graph $G$ over the sub-compression domains, where each node in $G$ stores the corresponding sub-compression domain together with the packs constructed during packing. The graph $G$ is then used for subsequent alignment-aware expansion.
    
    \item Third, we perform a topological traversal of $G$, iteratively constructing initial packs and expanding them. Specifically, starting from a maximal feasible group, packs are expanded along two levels of data dependencies. First, the algorithm follows dependencies in $G$ to identify the target compression domain $S.domain$. 
    Then, based on producer–consumer relationships, it selects nodes within $S.domain$ that consume the data produced by the seed pack $P_F$ and groups them into a new pack $P_S$.
    Once a new pack is formed, the associated node $S$ is updated by removing the packed nodes from $S.domain$ and inserting the new pack into $S.packs$.
    
    \item Finally, intra-domain merging is applied to improve bit-space utilization once the alignment-aware expansion is finished.
\end{itemize}

Overall, the algorithm follows an “alignment-first, density-second” idea, mitigating the conflict between bit-space utilization and data reuse efficiency, and achieves high utilization with limited data rearrangement overhead.

\subsection{Bit-level Compression Execution}\label{sec:compression_execution}

\begin{figure}[htbp]
    \SetCaptionSpacing{0.10cm}{0.10cm}
    \centering
    \includegraphics[width=\linewidth]{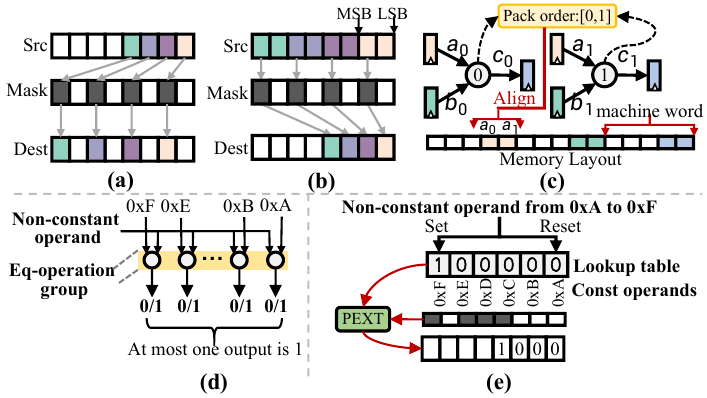}
    \caption{
            Examples of (a) \texttt{PDEP} and (b) \texttt{PEXT};
            (c) Pack-aligned memory layout;
            (d) An Eq-operation group;
            (e) LUT-based compression execution.
    }
    \label{fig:BMI}
\end{figure}

Operations in RTL graphs, referred to as \textit{RTL operations}, exhibit diverse dependency patterns across \subwordTerm s, complicating bit-level parallel execution. 
In general, these RTL operations can be categorized into two types. 
\textit{\subwordTermBegin-independent operations} allow each slice to be computed independently without cross-slice interference, thus enabling direct bit-level parallelization. 
In contrast, \textit{cross-\subwordTerm~dependent operations} are coupled across slices or have variable slice widths, preventing naive bit-level parallel execution. 

To address this challenge, we categorize RTL operations based on their dependency characteristics and adopt tailored bit-level compression execution strategies that preserve correctness while enabling parallel execution wherever possible. 
In this section, we present a unified bit-level compression execution scheme that efficiently supports both types of RTL operations.

\subsubsection{\textbf{Bit Manipulation Instructions}}
Before describing our bit-level compression execution, we introduce two key instructions from the Bit Manipulation Instructions~(BMI) set~\selfcite{TC_2009_Hilewitz}: \texttt{PDEP} and \texttt{PEXT}, which allow efficient bit-level deposition and extraction:
\begin{itemize}
     
    \item \texttt{\textbf{PDEP}} (Parallel Bit Deposit) deposits bits from a source word into a destination word according to a mask, as shown in Figure~\selfref{fig:BMI}~(a).

    \item \texttt{\textbf{PEXT}} (Parallel Bit Extract) extracts bits from a source word according to a mask and packs them contiguously into the least significant bits of the destination word, as shown in Figure~\selfref{fig:BMI}~(b).
\end{itemize}

\subsubsection{\textbf{Compression Execution for Logical Operations}}
Logical operations, such as \texttt{AND}, \texttt{OR}, and \texttt{NOT}, form the majority of RTL operations (over 55\%, shown in Figure~\selfref{fig:motivation1}~(b)) and naturally satisfy the \subwordTerm-independence property, since they do not propagate carries across bits. 
Consequently, multiple \subwordTerm s can be executed in parallel without introducing any data dependencies.
This provides a straightforward and efficient bit-level compression execution.

\subsubsection{\textbf{Compression Execution for Memory Access Operations}}
Memory access operations, primarily originating from registers, also satisfy the \subwordTerm-independence property, allowing multiple \subwordTerm s to access memory in parallel.
Leveraging this characteristic, we adopt a \textit{pack-aligned memory layout} shown in Figure~\selfref{fig:BMI}~(c), where each pack exclusively occupies a single machine word with its data layout aligned to the node order within the pack.
This memory layout avoids separate memory access and data rearrangements for individual \subwordTerm s, reducing both access and rearrangement overhead.

\subsubsection{\textbf{Compression Execution for Eq Operations}}
Eq operations inherently exhibit cross-\subwordTerm{} dependencies, meaning that the result of an individual \subwordTerm{} cannot be obtained independently. 
Nonetheless, we observe distinct opportunities for compression execution across different design types.

For AI accelerators, quantization techniques~\selfcite{quantization_acc1,quantization_acc2,quantization_acc3,quantization_acc4} are widely adopted, resulting in low bit-width operands.
Leveraging this characteristic, Eq operations between two packed operands $X$ and $Y$ can alternatively be implemented by first computing the \texttt{XOR} and then using \texttt{PEXT} combined with bitwise \texttt{OR} operations. 
Specifically, let $M_i$ denote the mask selecting the $i$-th bit from all \subwordTerm s. 
We first compute the \texttt{XOR} of the two packed operands, 
$
Z = X \oplus Y
$, which identifies differing bits. 
Then, for each bit position, we extract bits from \subwordTerm s using \texttt{PEXT}, 
$
Z_i = \texttt{PEXT}(Z, M_i)
$, 
and finally, the equality result for all \subwordTerm s is obtained by 
$
R = \neg \bigvee_{i=0}^{k-1} Z_i
$, 
where $k$ is the \subwordTerm{} bit-width.
Note that this approach is beneficial only for Eq operations with small operand bit-widths~(e.g., $\leq 4$).

For CPU designs, operands are typically wider, rendering the above approach ineffective. 
However, we observe that many Eq operations form groups in which one operand is a shared non-constant operand whose value is produced at runtime, while the other operand is a compile-time constant literal.
An example is shown in Figure~\selfref{fig:BMI}~(d).
For each Eq-operation group, the non-constant operand takes a fixed value in a simulation cycle, so at most one Eq operation in the group evaluates to 1 while the others are 0.  
To exploit this property, we adopt a \textit{lookup table (LUT)-based compression execution method} shown in Figure~\selfref{fig:BMI}~(e).
The LUT size is determined by the bitwidth of the operands, whose value range is known at compile time.
In this method, each LUT entry stores a Boolean value indicating the result of an Eq operation with a particular constant operand. 
The constant operand determines which entry in the LUT is associated, while the non-constant operand acts as an index to update the entry. 
Therefore, the execution of each Eq-operation group is transformed into a LUT update: 
at the start of a simulation cycle, the entry indexed by the non-constant value is reset, and the other entry indexed by the newly computed non-constant value is subsequently set.
This approach efficiently compresses multiple Eq operations into a single LUT update.
Furthermore, \texttt{PEXT} can also be applied to extract bits from the LUT in parallel.
Besides, to avoid excessive memory overhead, we limit the LUT size to 512 entries, which requires only 64 bytes of storage and is therefore negligible.

\subsubsection{\textbf{Compression Execution for Special Operations}}
Certain special operations, including shift-based operations and selective bit extraction, also exhibit cross-\subwordTerm{} dependencies. 

\textbf{Shift Operations:} Shift-based operations, such as left shifts, typically involve a constant operand in RTL simulation. 
At compile time, we construct a mask according to the \subwordTerm{} layout and constant operand, and at runtime, \texttt{PDEP} deposits the source bits into their target positions. 
Figure~\selfref{fig:BMI}~(a) illustrates an example where \texttt{PDEP} performs a left shift by one bit across all \subwordTerm s.

\textbf{Bit Extraction Operations:} These operations extract selected bits from \subwordTerm s, usually specified by a constant operand. 
We implement them using \texttt{PEXT}, generating the mask at compile time and extracting bits at runtime. 
Figure~\selfref{fig:BMI}~(b) shows an example where the most significant bit of each \subwordTerm{} is extracted.

\section{Evaluation}

\subsection{Setup}\label{section:setup}
The experiments are conducted on a server equipped with two Intel Xeon Platinum 8338 processors, each featuring 64 cores. Each core is configured with a private 32KB L1 instruction cache, a private 48KB L1 data cache, and a private 1.25MB L2 cache. 
Each processor includes a shared 48MB L3 cache.
\textit{Clang++-12} is used as the default compiler throughout this study, with optimization level {-O2}.

\Revision{
The simulators used for comparison include Verilator~(version 5.036)~\selfcite{verilator}, Arcilator~\selfcite{circt}, Khronos~\selfcite{zhou2023khronos}, Repcut~\selfcite{repcut}, and RTeAAL Sim~\selfcite{RTeAAL_ASPLOS_2026} as cycle-based compiled RTL simulation baselines aligned with ARMOR.
Dedup~\selfcite{dedup} and ESSENT~\selfcite{essent} are also included, although they are event-driven, because Dedup also targets front-end bottlenecks in RTL simulation through code reuse, and ESSENT is its baseline.
In addition, FIRRTL IR is adopted as the canonical input representation, as it can be converted to other input formats required by different simulators, such as CIRCT IR~\selfcite{circt} and Verilog.
Notably, RTeAAL Sim uses a loop-based representation that enables several loop-unrolling optimizations; we therefore adopt its PSU configuration, which provides the best performance.
}{A}

\subsection{Benchmark Evaluation}
We evaluate two representative design categories: AI accelerators and CPU designs. 
The designs are generated using the TensorLib~\selfcite{tensorlib} and Chipyard~\selfcite{chipyard}. 
Both frameworks support scalable hardware configurations by varying the number of processing elements~(PEs) or cores.
We select the following designs as benchmarks~(Table~\selfref{benchmark}):
\begin{itemize}
    \item \textbf{GEMM}-\textit{n}$\times$\textit{n}: 
    General Matrix-Matrix Multiplication~(GEMM) accelerator with an \textit{n}$\times$\textit{n} PE array, where each PE operates on 8-bit operands.
    
    \item \textbf{Conv2D}-\textit{n}$\times$\textit{n}: 
    Convolution accelerator with an \textit{n}$\times$\textit{n} PE array, where each PE operates on 8-bit operands.

    \Revision{
    \item \textbf{Gemmini}-\textit{n}$\times$\textit{n}: 
    DNN accelerator with an \textit{n}$\times$\textit{n} PE array, where each PE operates on 8-bit operands~\selfcite{gemmini-dac}.
    }{A}
    
    \item \textbf{Rocket}-\textit{n}\textbf{C}: RocketChip with \textit{n} in-order RISC-V Medium Rocket cores~\selfcite{rocket_core}. 
    
    \item \textbf{Boom}-\textit{n}\textbf{C}: Boom with \textit{n} out-of-order RISC-V Small Boom cores~\selfcite{boom_core}.
    
    \Revision{
    \item \textbf{Boom}-\textbf{M}\textit{n}\textbf{C}: Boom with \textit{n} out-of-order RISC-V Mega Boom cores~\selfcite{boom_core}.
    }{A}
\end{itemize}

\begin{table}[h]
\centering
\SetCaptionSpacing{0.10cm}{0.10cm}
\caption{Evaluation of benchmark scale and module reuse.}
\label{benchmark}
\resizebox{\linewidth}{!}{
\begin{tabular}{l|r|r|r|r|r}  
\hline
    \multirow{1}{*}{\textbf{Benchmark}} & 
    \multirow{1}{*}{\textbf{IR Nodes}} & 
    \multirow{1}{*}{\textbf{IR Edges}} & 
    \textbf{Module} & 
    \textbf{Instance} & 
    \textbf{Reuse} \\
\hline
\hline
GEMM-16$\times$16 & 21410 & 30345 & 14 & 1778 & 99\% \\
Conv2D-16$\times$16 & 93350 & 150838 & 21 & 1570 & 99\% \\
Gemmini-32$\times$32 & 154948 & 261205 & 99 & 3240 & 97\% \\
Rocket-8C & 122062 & 202753 & 341 & 1297 & 74\% \\
Boom-2C & 169628 & 308586 & 461 & 1098 & 58\% \\
Boom-8C & 611397 & 1123760 & 461 & 2745 & 83\% \\
Boom-M8C & 2862097 & 5452476 & 378 & 4365 & 91\% \\
\hline
Average & - & - & - & - & 85\%\\
\hline
\end{tabular}
}
\end{table}
Table~\selfref{benchmark} summarizes the characteristics of the evaluated benchmarks, including the scale of RTL graphs, the number of modules and instances, and the module reuse ratio. 
On average, the module reuse ratio reaches 85\%, indicating a high degree of structural reuse in the highly parallel designs. 
Such structural regularity implies the presence of abundant isomorphic subgraphs in the RTL graphs, thereby creating substantial opportunities for node compression.
\begin{figure}[htbp]
    \SetCaptionSpacing{0.10cm}{0.10cm}
    \centering
    \includegraphics[width=0.8\linewidth]{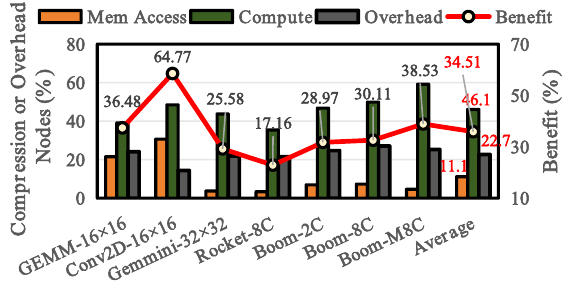}
    \caption{
        Breakdown of node compression, including memory access and computation nodes, together with the overhead of data rearrangement, and the resulting net reduction. 
        The average values are highlighted in red.
    }
    \label{fig:compression_rate}
\end{figure}

Figure~\selfref{fig:compression_rate} provides a detailed breakdown of node compression along with the overhead introduced by data rearrangement. The overhead is quantified by the number of introduced data packing and unpacking operations.
On average, 11.1\% of memory access nodes and 46.1\% of computation nodes are compressed. This compression comes at a cost of 22.7\% overhead from data rearrangement. 
Nevertheless, after accounting for this overhead, the proposed node compression still achieves an average net reduction of 34.51\% in the size of the RTL graph.

\begin{figure}
    \SetCaptionSpacing{0.10cm}{0.10cm}
    \centering
    \includegraphics[width=\linewidth]{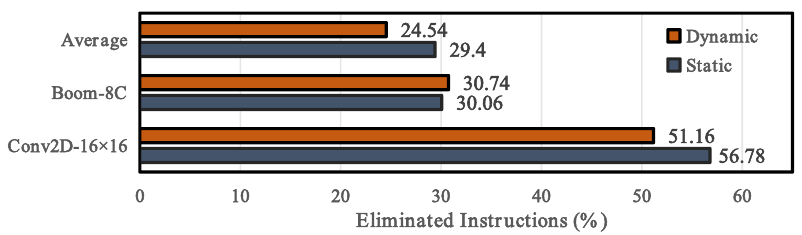}
    \caption{
        \Revision{Percentage of static and dynamic instructions eliminated with node compression.}{D}
    }
    \label{fig:instruction_eliminated}
\end{figure}

\begin{figure*}
    \SetCaptionSpacing{0.10cm}{0.10cm}
    \centering
    \includegraphics[width=1\linewidth]{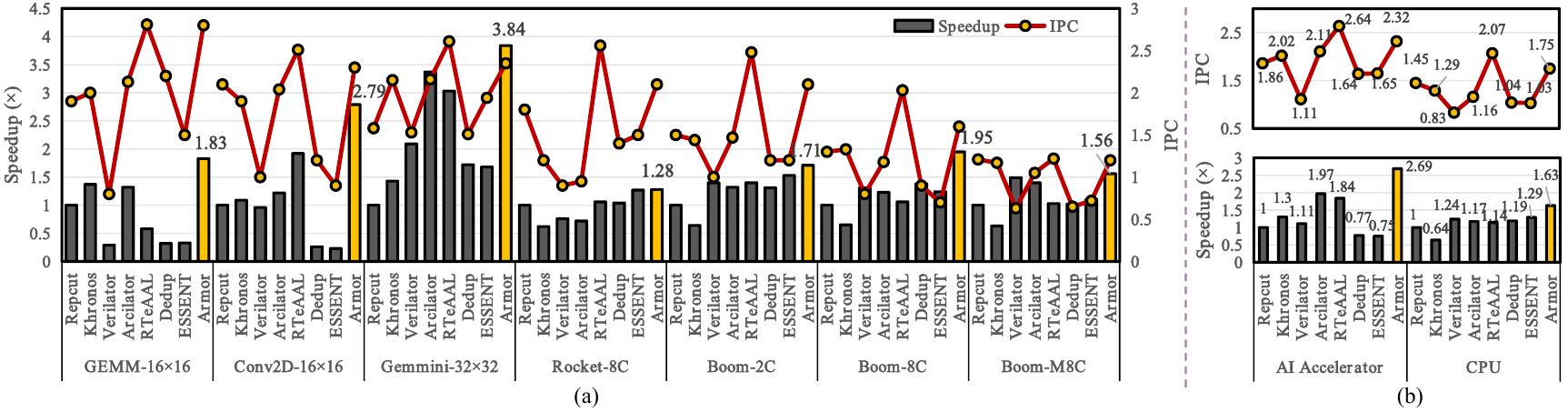}
    \caption{
        (a) Comparison of simulation performance in terms of speedup and IPC.  
        (b) Average speedup and IPC across AI accelerator and CPU designs.  
        The baseline is Repcut.
        \papertitle{} is highlighted with yellow bars, and its speedup values are annotated on the bars. 
    }
    \label{fig:runtime}
\end{figure*}

Figure~\selfref{fig:instruction_eliminated} quantifies the implicit code reuse enabled by node compression. 
Unlike conventional function-level code reuse, node compression reuses code implicitly at the instruction-sequence level: one compressed word-level instruction sequence represents the scalar execution of multiple isomorphic RTL nodes before compression. 
We quantify the code reuse enabled by node compression using the eliminated instruction percentage.
From the static-code perspective, node compression achieves an average static implicit code reuse rate of 29.4\%, meaning that 29.4\% of the generated instruction footprint is removed by reusing one compressed instruction sequence for multiple isomorphic nodes. 
From the dynamic-execution perspective, it achieves an average dynamic implicit code reuse rate of 24.54\%, meaning that 24.54\% of dynamically executed scalar instructions are avoided through compressed execution.



\subsection{Performance Comparison}

\subsubsection{\textbf{Runtime comparison}}

We evaluate the performance of various simulators across all the aforementioned benchmarks.
Figure~\selfref{fig:runtime} presents the speedup of each simulator relative to Repcut, along with its instructions per cycle~(IPC).
As shown in Figure~\selfref{fig:runtime}, \papertitle{} consistently achieves the best performance for both AI accelerator and CPU designs. Specifically, \papertitle{} attains a peak speedup of 3.84$\times$ on Gemmini-$32\times32$ and 1.95$\times$ on Boom-8C.
On average, \papertitle{} achieves speedups of 2.69$\times$ on AI accelerator designs and 1.63$\times$ on CPU designs.  
In terms of IPC, \papertitle{} also outperforms other simulators, except for RTeAAL.
Specifically, \papertitle{} achieves average IPCs of 2.32 on AI accelerator designs and 1.75 on CPU designs, both substantially higher than Verilator, which shows the lowest IPCs of 1.11 and 0.83, respectively.
The higher IPC achieved by \papertitle{} can be attributed to its effective mitigation of front-end bottlenecks, preventing pipeline stalls caused by insufficient instruction supply.
A detailed analysis of this effect is presented in Section~\selfref{sec:frontend_cmp}.



Interestingly, for the event-driven simulators ESSENT and Dedup, simulation performance on AI accelerator designs is consistently inferior to that of cycle-based simulators. 
In contrast, for CPU designs, they generally outperform cycle-based simulators, except in the case of \papertitle{}.
This phenomenon is determined by the intrinsic signal activity patterns of the two types of designs. 
The high signal activity in AI accelerator designs~\selfcite{synopsys_pcie_ai_2026,high_power_ai} makes the overhead of event scheduling and management in event-driven simulation exceed the cost of redundant computation in cycle-based simulation. 
Conversely, CPU designs exhibit extremely low signal activity~\selfcite{essent,low_activity_1,dark_silicon}, enabling event-driven simulation to prune inactive computations and thus achieve higher performance.
Despite the above observations, \papertitle{} consistently achieves the best performance among all simulators, demonstrating its robustness across diverse design types.
It is worth noting that although \papertitle{} is a cycle-based simulator, the node compression technique proposed in this paper is not limited to cycle-based simulation and can also be applied in event-driven simulation.

\subsubsection{\textbf{Front-End Bottleneck Comparison}}\label{sec:frontend_cmp}

To evaluate \papertitle{}'s effectiveness in mitigating front-end bottlenecks, we conduct a detailed micro-architectural analysis using both Perf~\selfcite{perf} and Intel VTune~\selfcite{intel_vtune}. 
VTune’s Top-Down Microarchitecture Analysis identifies performance bottlenecks of software running on Intel processors, focusing on four metrics: Retiring, Front-end bound, Back-end bound, and Bad speculation~\selfcite{frontend_topdown}.

\begin{figure}
    \SetCaptionSpacing{0.10cm}{0.10cm}
    \centering
    \includegraphics[width=\linewidth]{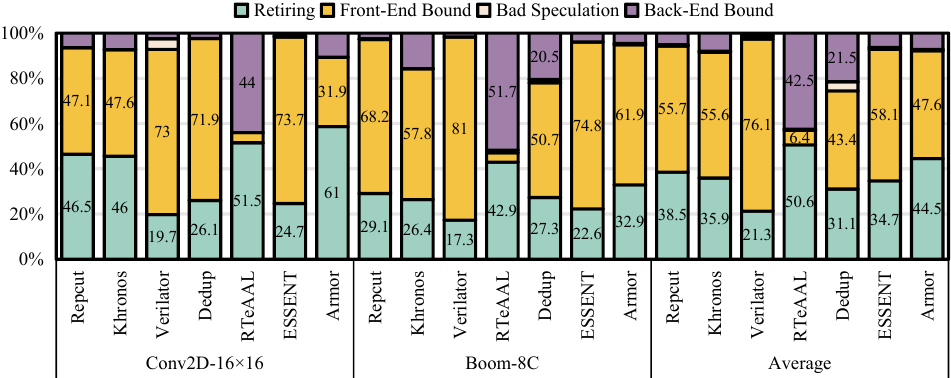}
    \caption{
    Micro-architecture results of Conv2D-$16\times16$, Boom-8C designs, and the average.
    }
    \label{fig:vtune}
\end{figure}

Figure~\selfref{fig:vtune} reports the top-down metrics for two designs, Conv2D-$16\times16$ (AI accelerator) and Boom-8C (CPU), as well as the averages across all evaluated benchmarks.
As expected, fully unrolled simulators suffer from more severe front-end bottlenecks than non-fully-unrolled simulators such as Dedup and RTeAAL Sim.
However, reducing full unrolling also introduces significant back-end pressure.
For example, the back-end bounds of RTeAAL Sim and Dedup reach 42.5\% and 21.5\%, respectively, which are much higher than those of fully unrolled simulators.
We do not further explore this front-end/back-end trade-off in this paper; instead, the following analysis focuses on the setting targeted by \papertitle{}: mitigating front-end bottlenecks in fully unrolled simulators.

\begin{figure}
    \SetCaptionSpacing{0.10cm}{0.10cm}
    \centering
    \includegraphics[width=\linewidth]{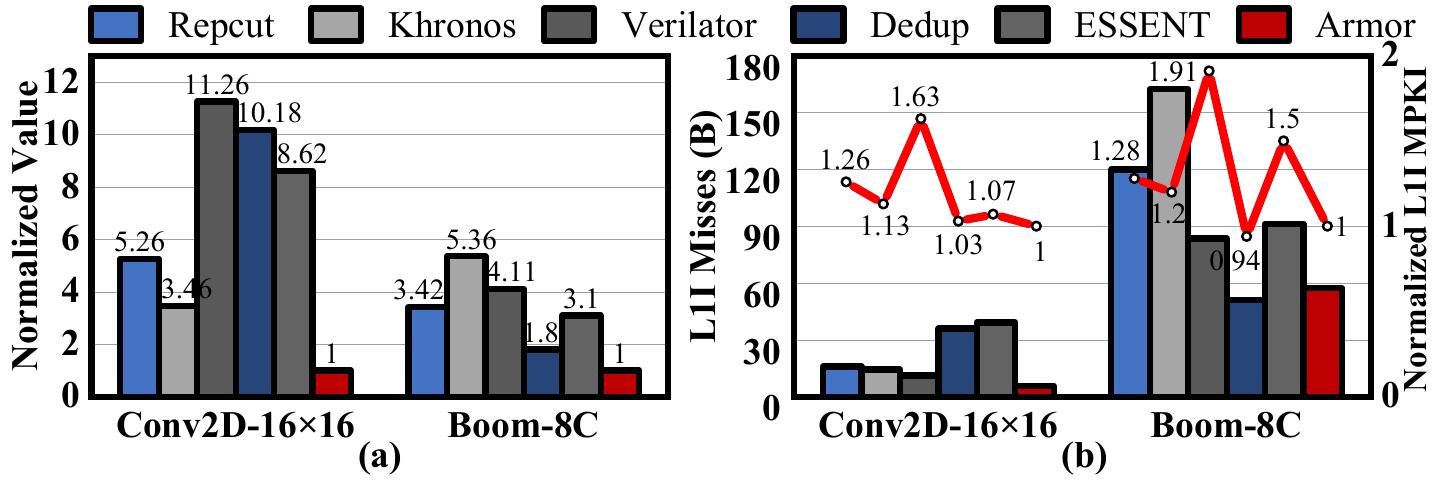}
    \caption{
        (a): Normalized \texttt{idq\_uops\_not\_delivered.core}.
        (b): L1I cache misses and normalized \texttt{L1I MPKI}.
        All normalized values are relative to \papertitle{}; lower is better.
    }
    \label{fig:perf_uop_icache}
\end{figure}

Among the fully unrolled simulators, \papertitle{} exhibits the lowest front-end bound for both AI accelerator and CPU designs, with an average front-end bound of 47.6\%, lower than Verilator's 76.1\%, the highest among the evaluated fully unrolled simulators.
Furthermore, we note that the front-end bound is calculated relative to each simulator’s total CPU execution cycles\footnote{Front-end bound = idq\_uops\_not\_delivered.core / (n $\times$ cycles), where $n$ is the CPU issue width.} rather than an absolute cross-simulator comparison. 
To obtain a more simulator-independent measure, we record \texttt{idq\_uops\_not\_delivered.core} events using Perf, which counts the number of micro-ops the front-end fails to deliver. 
As shown in Figure~\selfref{fig:perf_uop_icache}~(a), \papertitle{} consistently achieves the lowest value of this metric among all evaluated simulators.
This indicates that \papertitle{} delivers micro-ops more efficiently, experiencing lower front-end pressure than other simulators.


Complementing this, Figure~\selfref{fig:perf_uop_icache}~(b) reports L1 instruction cache misses, and L1 instruction cache misses per kilo instructions~(L1I MPKI). 
Overall, \papertitle{} exhibits fewer L1 cache misses and lower L1I MPKI than other simulators. 
This reduction directly decreases instruction fetch stalls and prevents pipeline stalls due to cache fills.
Consequently, the front-end can supply instructions to the back-end more steadily, mitigating instruction starvation.
Taken together, these analyses show that \papertitle{} improves both instruction fetch efficiency and front-end instruction supply, thereby effectively alleviating front-end bottlenecks and ultimately enhancing overall CPU pipeline utilization compared to other simulators.

\subsubsection{\textbf{Bit-space utilization comparison}}\label{sec:bit_utilization}
\begin{figure}
    \SetCaptionSpacing{0.10cm}{0.10cm}
    \centering
    \includegraphics[width=\linewidth]{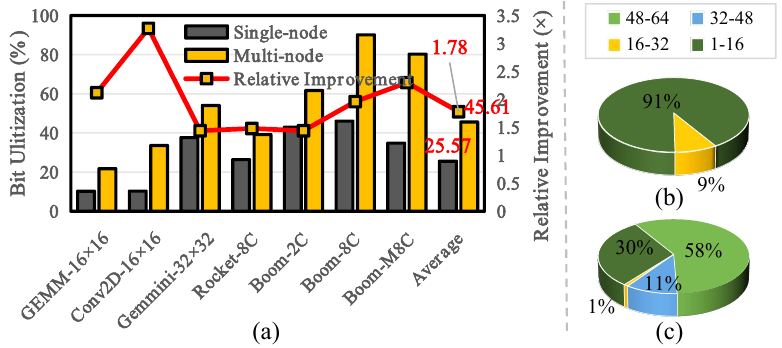}
    \caption{
    (a) Bit-space utilization comparison between single-node execution and multi-node execution.
    Node bit-width distribution in the Conv2D-$16\times16$ design under the single-node~(b) and multi-node~(c) execution.
    }
    \label{fig:bit_utilization}
\end{figure}
To evaluate the improvement in machine-word utilization, we compare the bit-space utilization between the single-node execution and our multi-node execution model.
As shown in Figure~\selfref{fig:bit_utilization}~(a), on average, the bit utilization of single-node execution is 25.57\%, while multi-node execution achieves 45.61\%, yielding a 1.78$\times$ improvement.
These results demonstrate that single-node execution leads to substantial under-utilization of machine words, whereas multi-node execution effectively improves bit-space utilization.

To better understand the source of the improvement in bit-space utilization, we analyze the distribution of node bit-widths in the Conv2D-$16\times16$ design, which exhibits the largest improvement in bit utilization across all designs. 
Node widths are grouped into four ranges: 
[1,16), [16,32), [32,48), and [48,64].
As shown in Figure~\selfref{fig:bit_utilization}~(b) under single-node execution, about 91\% of nodes have bit-widths smaller than 16 bits, leading to substantial waste of machine-word space. 
In contrast, with our node compression, the utilized bit-widths of approximately 70\% of the packs exceed 32 bits, as shown in Figure~\selfref{fig:bit_utilization}~(c), significantly improving machine-word utilization.

\subsection{Impact of Architectural Parallelism Degree}
\begin{figure}
    \SetCaptionSpacing{0.10cm}{0.10cm}
    \centering
    \includegraphics[width=\linewidth]{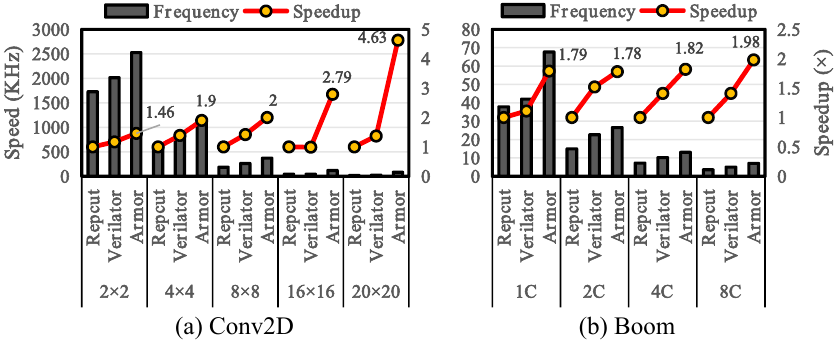}
    \caption{
        The performance of RTL simulators and the corresponding speedup relative to Repcut across different architectural parallelism configurations.
    }
    \label{fig:parallelism}
\end{figure}

To evaluate how simulation performance changes with the degree of architectural parallelism, we scale the number of replicated modules in the RTL design. 
Specifically, we select benchmarks from both AI accelerators and CPU designs, varying parallelism by scaling the PE array size for Conv2D ($2\times2$ to $20\times20$) and the number of cores for Boom (1 to 8).

As shown in Figure~\selfref{fig:parallelism}, the performance advantage of \papertitle{} becomes increasingly pronounced as the degree of parallelism grows. 
For instance, for the Conv2D design, the speedup over Repcut improves from 1.46$\times$ under the 2$\times$2 configuration to 4.63$\times$ at 20$\times$20. 
Although the performance gain for Boom is less pronounced, it still shows a steady increasing trend with higher parallelism.
This trend is attributed to higher architectural parallelism, which introduces more substantial isomorphic structures in the RTL graph, providing greater opportunities for node compression and ultimately more effectively alleviating front-end bottlenecks.
Consequently, \papertitle{} is able to achieve greater performance gains.
More importantly, this trend further highlights that \papertitle{} is expected to exhibit even more pronounced performance advantages on future designs with higher parallelism~\selfcite{HPCA_2026_Yu}.
Conversely, these results also indicate that designs with lower module reuse or more irregular structures provide fewer compression opportunities, so the benefit of node compression is expected to be smaller.
Overall, these results demonstrate that \papertitle{} achieves superior scalability by effectively leveraging structural isomorphism introduced by highly parallel architectures, enabling consistently higher performance gains across both AI accelerator and CPU designs.


\subsection{Impact of Compression Granularity}\label{sec:granluarity}
\begin{figure}
    \SetCaptionSpacing{0.10cm}{0.10cm}
    \centering
    \includegraphics[width=0.9\linewidth]{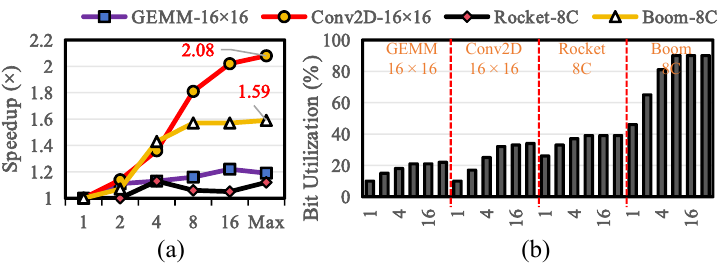}
    \caption{
        Speedup (a) and Bit-space utilization (b) of \papertitle{} under different compression granularities.
        The speedup is normalized to the compression granularity of 1.
    }
    \label{fig:granluarity}
\end{figure}

To analyze the impact of the number of compressed nodes per machine word on simulation performance, we vary the compression granularity by limiting the maximum number of nodes packed into a single machine word. 
Specifically, we limit the number of compressed nodes to 1, 2, 4, 8, 16, and \textit{Max}, where \textit{Max} represents the hardware-imposed upper bound on the pack size.
Figure~\selfref{fig:granluarity} shows the impact of compression granularity on simulation performance and bit-space utilization. 
As shown in Figure~\selfref{fig:granluarity}~(a), increasing the number of compressed nodes per machine word improves simulation performance for both AI accelerator and CPU designs. 
This is because larger pack sizes enable more effective exploitation of bit-level parallelism, reducing both the number of executed operations and the code footprint, thereby improving simulation performance.
Correspondingly, packing more nodes into one machine word also increases bit-space utilization, as illustrated in Figure~\selfref{fig:granluarity}~(b).


\subsection{Impact of Data width}
\begin{figure}
    \SetCaptionSpacing{0.10cm}{0.10cm}
    \centering
    \includegraphics[width=0.9\linewidth]{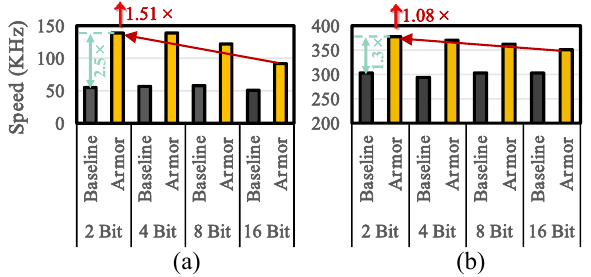}
    \caption{
        RTL simulation performance under different data widths of AI accelerators: (a) Conv2D-$16\times16$; (b) GEMM-$16\times16$.
    }
    \label{fig:data_width}
\end{figure}
Modern AI accelerators widely adopt quantization techniques, such as INT8, and even lower-precision data representations, to reduce computation and memory cost~\selfcite{quantization_acc1,quantization_acc2,quantization_acc3,quantization_acc4}. 
Since data width affects the number of nodes that can be packed into a single machine word in \papertitle{}, we evaluate its impact on the performance of \papertitle{} across different data widths.
In this experiment, we evaluate widely adopted quantization data widths of 2-bit, 4-bit, 8-bit, and 16-bit on the Conv2D-$16\times16$ and GEMM-$16\times16$ designs, while keeping the architecture unchanged.
Figure~\selfref{fig:data_width} shows the performance of the baseline (without node compression) and \papertitle{} across different data widths. 
We observe that the baseline remains insensitive to the data width, whereas \papertitle{} benefits from reduced data widths. 
For example, for the Conv2D-$16\times16$ design, \papertitle{} achieves a 1.51$\times$ speedup when the data width decreases from 16 bits to 2 bits, and delivers up to 2.5$\times$ speedup over the baseline at 2 bits.
This improvement is driven by similar factors as discussed in Section~\selfref{sec:granluarity}.
As the data width decreases, more nodes can be packed into a single machine word, 
thereby improving performance.


\subsection{Impact of Cache Capacity}
\begin{figure}
    \SetCaptionSpacing{0.10cm}{0.10cm}
    \centering
    \includegraphics[width=0.9\linewidth]{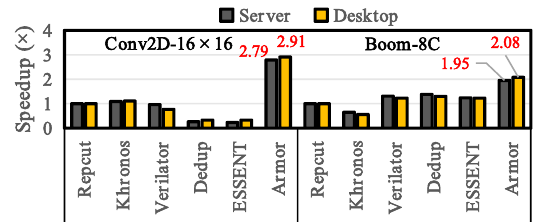}
    \caption{
        Performance of different RTL simulators on server and desktop platforms, normalized to Repcut.
        The speedup of \papertitle{} is highlighted in red.
    }
    \label{fig:cache}
\end{figure}
To evaluate the impact of cache capacity, we conduct experiments on a desktop platform with a smaller cache hierarchy (L1I: 32KB, L2: 256KB, L3: 16MB), compared to the server platform described in Section~\selfref{section:setup}.
To avoid environmental differences~(e.g., CPU frequency) between the two platforms, all results are normalized to Repcut instead of absolute performance.
Figure~\selfref{fig:cache} shows the performance of different RTL simulators on the server and desktop platforms.
We can obtain two observations.
First, across both the server and desktop platforms, \papertitle{} consistently outperforms other simulators, demonstrating its robust performance and strong generality across different hardware platforms.
Second, unlike other simulators, \papertitle{} achieves higher speedups on the desktop platform than on the server.
This improvement stems from its effective alleviation of cache pressure, making \papertitle{} more cache-friendly and particularly effective under constrained cache capacities, further highlighting \papertitle{}'s ability to alleviate front-end bottlenecks.

\begin{figure}
    \SetCaptionSpacing{0.10cm}{0.10cm}
    \centering
    \includegraphics[width=\linewidth]{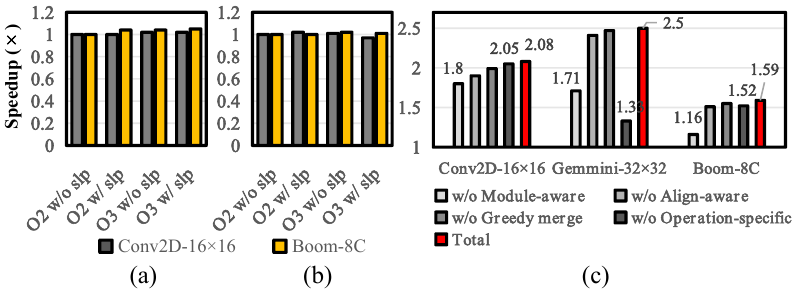}
    \caption{
        Performance of Verilator (a) and ARMOR (b) under four compiler configurations, with speedups normalized to \textit{O2 w/o SLP}.
        (c) ARMOR performance when removing each component, with speedups normalized to the configuration without node compression.
    }
    \label{fig:ablation}
\end{figure}


\subsection{Ablation Study}

\subsubsection{\textbf{Ablation of C++ Compiler}}

To evaluate the impact of compiler optimization and SLP vectorization, we compile both ARMOR-generated and Verilator-generated simulators under four configurations: \textit{O2~w/o~SLP}, \textit{O2~w/~SLP}, \textit{O3~w/o~SLP}, and \textit{O3~w/~SLP}.
As shown in Figure~\selfref{fig:ablation}~(a) and~(b), enabling SLP vectorization or increasing the optimization level provides only limited performance improvement for both simulators, with gains below 5\%, indicating that compiler auto-vectorization cannot effectively recover the compression opportunities exploited by ARMOR.

\subsubsection{\textbf{Ablation of \papertitle{} Components}}

To estimate the contribution of each component, we evaluate the following variants of ARMOR.
\begin{itemize}
    \item \textit{\textbf{w/o Module-aware}} replaces ARMOR's module-aware pack discovery with a seed-driven search inspired by SLP~\selfcite{PLDI_2000_Larsen, PACT_2017_SuperGraph}: starting from register nodes in the RTL graph, it searches candidate packs along \textit{use}-\textit{def} and \textit{def}-\textit{use} chains.

    \item \textit{\textbf{w/o Align-aware}} keeps aligned packing inside each module but disables alignment-aware packing across module boundaries, which increases data rearrangement at module boundaries.

    \item \textit{\textbf{w/o Greedy merge}} preserves the initial dense pack using \textsc{SelectMaxFeasibleGroup} but disables the subsequent intra-domain pack merging.

    \item \textit{\textbf{w/o Operation-specific}} applies node compression only to slice-independent operations, while cross-slice-dependent operations remain uncompressed and use the original per-node execution.

    \item \textit{\textbf{Total}} denotes the full ARMOR design.
\end{itemize}

As shown in Figure~\selfref{fig:ablation}~(c), each component contributes to the performance of \papertitle.
As expected, module-aware pack discovery shows the largest impact because it provides a subgraph-level view for pack discovery; without this view, local seed-based search is more likely to make short-sighted packing decisions that increase data rearrangement overhead.
We also observe that \textit{w/o Operation-specific} causes the largest slowdown on Gemmini-$32\times32$. Further analysis shows that Gemmini-$32\times32$ contains many bit-extraction operations, which frequently break data reuse and increase data rearrangement overhead when operation-specific compressed execution is disabled.
Overall, the full ARMOR design achieves the best performance, showing that each component is necessary for profitable node compression.

\section{Discussion \& Future Work}


\Revision{
\textbf{Relation to SIMD:}
ARMOR currently uses scalar machine words as the execution substrate, packing RTL nodes into software-defined bit slices to improve bit-space utilization.
Rather than directly mapping RTL nodes to SIMD lanes, a natural integration is first to apply ARMOR's bit-level node compression within each word, and then use SIMD to process multiple compressed words in parallel.
}{D}




\Revision{

\textbf{Portability of BMI2:} 
ARMOR currently uses BMI2 \texttt{PDEP}/\texttt{PEXT} instructions to implement cross-slice dependent operations after node compression. 
On platforms where BMI2 \texttt{PDEP}/\texttt{PEXT} is unavailable, ARMOR falls back to a conservative mode by disabling compression for cross-slice dependent operations and only compressing slice-independent operations. 
}{B}

\Revision{
\textbf{Complementarity with Parallel Simulation:} 
ARMOR is orthogonal to partitioning-based multi-threaded RTL simulation, such as RepCut-style~\selfcite{repcut} approaches.
A practical integration is to first partition the RTL graph across threads and then apply node compression locally within each partition.
Since compressed packs are formed inside a partition, ARMOR does not introduce additional cross-thread dependencies or synchronization, and it preserves partition-local code and data locality.
The main open issue is load balancing: different partitions may expose different compression opportunities, so their post-compression workloads may deviate from the original partitioning cost model.
}{A, B}


\textbf{Future Work:} 
An important direction for future work is to investigate the trade-off between scalar and SIMD execution. Some cross-slice dependent operations may benefit more from SIMD execution, motivating the exploration of how compressed packs should be mapped to scalar or SIMD execution units based on their operation characteristics and data layouts.

\section{Related work}


The performance of RTL simulation has already become a bottleneck in hardware design~\selfcite{898828, gupta2020improving}. 
Prior research efforts can be categorized into the following types.


\textbf{Sequential RTL Simulation.} 
Verilator~\selfcite{verilator} is an open-source simulator, capable of generating fast, cycle-accurate models for digital circuits. 
ESSENT~\selfcite{essent} combines event-driven and full-cycle simulation by employing a coarsened, conditional, singular, static execution model that exploits low activity while minimizing event management overhead.
GSIM~\selfcite{GSIM} further builds on ESSENT by reducing event management overhead and redundant computations.
Dedup~\selfcite{dedup} proposes a coarse-grained circuit deduplication strategy, which identifies multiple instances of a single module within a digital circuit and creates shared code that can be applied to all of these instances. 
Khronos~\selfcite{zhou2023khronos} optimizes RTL simulation by fusing inter-cycle memory accesses, which performs well for domain-specific accelerators employing deep pipelines. 
RTeAAL Sim~\selfcite{RTeAAL_ASPLOS_2026} reformulates RTL simulation as a sparse tensor algebra problem that can incorporate a wide range of tensor algebra optimizations.
Despite optimizing RTL simulation from different perspectives, many existing compiled simulators still rely on per-node lowering, while approaches such as Dedup and RTeAAL reduce code footprint through reusable or rolled representations at the cost of static specialization. 
In contrast, \papertitle{} compresses multiple RTL nodes into shared instruction sequences, reducing code footprint while preserving the static specialization of fully unrolled RTL simulation.

\textbf{Parallel RTL simulation.}
Recent research explores parallelization techniques to boost RTL simulation performance. 
RepCut~\selfcite{repcut} reduces inter-thread synchronization by replicating small overlaps, achieving a near-linear speedup.  
RTLFlow~\selfcite{lin2022rtl} is a GPU-accelerated RTL simulator that exploits stimulus-level parallelism.
GEM~\selfcite{GEM} introduces an emulator‑inspired methodology that compiles RTL into a GPU‑friendly virtual Very Long Instruction Word representation, enabling high‑performance RTL simulation on GPUs.
Parendi~\selfcite{parendi} exploits the abundant fine-grained parallelism and maps it onto the massively parallel Graphcore Intelligence Processing Unit to enable large-scale parallel RTL simulation.

\textbf{FPGA Emulation and Hardware Accelerator.}
FPGA prototypes map RTL circuits to FPGA gates for high-speed simulation, but face capacity and compile-time limitations.
FireSim~\selfcite{firesim} is an open-source FPGA emulation platform for cycle-accurate simulation, and FireAxe~\selfcite{FireAxe} extends it with user-guided partitioning across multiple FPGAs.
Commercial emulation platforms, such as Cadence Palladium~\selfcite{cadence_emulation}, employ dedicated emulation hardware to support high-speed execution of large-scale RTL designs.
Manticore~\selfcite{emami2023manticore} is an RTL simulation accelerator that uses a static bulk-synchronous parallel execution model to eliminate fine-grain synchronization overhead. 
ASH~\selfcite{elsabbagh2023accelerating} is a parallel RTL simulation architecture that combines fine-grained dataflow and selective event-driven execution to efficiently exploit both parallelism and low activity in RTL simulation.


\section{Conclusion}

This paper presents \papertitle{}, which alleviates front-end bottlenecks through node compression. 
\papertitle{} leverages bit-level data parallelism, allowing a single instruction sequence to serve multiple nodes instead of one per node, reducing the code footprint. 
\papertitle{} introduces a module-aware isomorphic subgraph identification method to mine compression opportunities, an alignment-aware dense node packing strategy to balance data reuse and bit-space utilization, and a unified bit-level parallelism scheme to support bit-level parallel execution of RTL operations. 
Experimental results show that \papertitle{} compresses, on average, 57\% of nodes, achieving speedups of 2.69$\times$ on AI accelerator designs and 1.63$\times$ on CPU designs, outperforming state-of-the-art simulators.

\section*{Acknowledgment}
We thank the anonymous reviewers for their insightful and constructive feedback. 
We also thank Shengwen Liang, Tianyun Ma, and Husheng Han for their valuable guidance and suggestions on the writing of this paper.
This work is supported by the Strategic Priority Research Program of the Chinese Academy of Sciences (XDB0660103) and the National Natural Science Foundation of China (92373206, U25A20486).
The corresponding authors are Jianan Mu, Zhiteng Chao, and Huawei Li.

\IEEEtriggeratref{57}
\bibliographystyle{IEEEtran}
\bibliography{sample-base}

\end{document}